\documentclass[reprint,twocolumn,superscriptaddress,showkeys,
nofootinbib,notitlepage,amsmath,amssymb,floatfix]{revtex4-2}

\pdfoutput=1
\usepackage{latexsym,amsmath,amssymb,lmodern,float,url}
\usepackage{natbib}
\usepackage{color}
\usepackage{multirow}
\usepackage{bm}
\usepackage{array}
\usepackage[normalem]{ulem}
\usepackage{cleveref}

\usepackage{mathtools}

\def\de{\delta^{\vphantom{1}}}
\def\bde{{\bar\delta}}

\def\ccss{{c\bar{c}s\bar{s}}}

\def\ccqq{{c\bar{c}q\bar{q}^\prime}}

\def\h3{{\displaystyle{\frac 3 2}}}

\newcommand{\bbar}{\overline}

\def\schro{Schr\"odinger~}

\begin{document}
\title{Exotic Hadrons from Scattering in the Diabatic Dynamical Diquark Model}
\author{Richard F. Lebed}
\email{Richard.Lebed@asu.edu}
\author{Steven R. Martinez}
\email{srmart16@asu.edu}
\affiliation{Department of Physics, Arizona State University, Tempe,
AZ 85287, USA}
\date{June, 2023}

\begin{abstract}
The diabatic framework generalizes the adiabatic approximation built into the Born-Oppenheimer (BO) formalism, and is devised to rigorously incorporate the mixing of BO-approximation eigenstates with two-particle thresholds.  We recently applied this framework in a bound-state approximation to the mixing of hidden-charm dynamical-diquark tetraquark states with open-charm di-meson thresholds.  Since almost all of these states are observed as above-threshold resonances, we here implement the corresponding scattering formalism to allow for a study of exotic tetraquark resonances within the diabatic framework.  We calculate elastic open-charm di-meson cross sections (in channels with zero, open, and hidden strangeness) as functions of center-of-mass energy, and observe the development of true resonances, near resonances, and various threshold cusp effects.  As an example, $\chi_{c1}(3872)$ can originate in the $1^{++}$ channel as a diquark-antidiquark state enhanced by the $D^0 \overline{D}^{*0}$ threshold, with or without an additional contribution from the conventional charmonium $\chi_{c1}(2P)$ state. 
\end{abstract}

\maketitle

\section{Introduction}

Reaching the 20-year anniversary of the first clear experimental evidence for the existence of heavy-quark exotic hadrons---the observation of the charmoniumlike state now called $\chi_{c1}(3872)$~\cite{Choi:2003ue}---the field of hadron spectroscopy now faces the same scientific challenges shared by many other areas of study.  Some definitive answers on the nature of these states have been obtained; but many of the original questions remain, and many new questions have arisen.  More than 60 heavy-quark exotic candidates have been observed to date, notably some of which that were first seen shortly after the 2003 discovery of $\chi_{c1} (3872)$ by Belle~\cite{Choi:2003ue} ({\it e.g.}, $Y(4260)$ in 2005 by BaBar~\cite{Aubert:2005rm}, which has subsequently been determined by BESIII to consist of more than one state~\cite{Ablikim:2016qzw}).  Despite a longstanding need for a theoretical paradigm to describe the structure, production, and decays of these states, no universally predictive model has emerged capable of accommodating all of them~\cite{Lebed:2016hpi,
Chen:2016qju,Hosaka:2016pey,Esposito:2016noz,Guo:2017jvc,Ali:2017jda,
Olsen:2017bmm,Karliner:2017qhf,Yuan:2018inv,Liu:2019zoy,
Brambilla:2019esw,Chen:2022asf,Lebed:2022vfu}.  A number of these exotic candidates (some of which are listed in Table~\ref{tab:exotics}) lie remarkably close to some particular di-hadron threshold, the most notable example being $\chi_{c1}(3872)$: 
\begin{equation}
m_{\chi_{c1}(3872)} - m_{D^0} - m_{D^{*0}} = -0.04 \pm 0.09 \ {\rm MeV} ,
\label{eq:XBE}
\end{equation}
using the averaged mass value for each particle provided by the Particle Data Group (PDG)~\cite{ParticleDataGroup:2022pth}.  Clearly, it can be no coincidence that so many of these states appear near a threshold.  Some of them, such as $\chi_{c1}(3872)$, lie close \textit{below} the corresponding threshold, suggesting a possible description via a di-hadron molecular picture, with the hadron pair (in this case, $D^0 \bar D^{*0}$ plus its charge-conjugate) being bound in part via $\pi^0$ exchange.  In fact, this interpretation has a rich history, in some cases long predating the $\chi_{c1}(3872)$ discovery~\cite{Voloshin:1976ap,DeRujula:1975qlm,Tornqvist:1993ng}.  Others, such as the $Z$ states of Table~\ref{tab:exotics}, lie close \textit{above} a threshold, discouraging the naive meson-exchange molecular description.  A complete, self-consisent model must be able to describe the relation between these exotic states and their nearby thresholds, as well as states that lie relatively far from any di-hadron threshold, such as $Z_c(4430)$ or many of the $J^{PC} = 1^{--}$ $Y$ states. 

Adding to the puzzle, $\chi_{c1}(3872)$ exhibits some behaviors that seem to imply the importance of short-distance components of its wavefunction, such as in its appreciable decays to $J/\psi$, $\chi_{c1}$, and $\gamma \psi(2S)$, with the radiative decays being especially significant in this regard.  However, given the tiny binding energy [Eq.~(\ref{eq:XBE})] available to a molecular $\chi_{c1}(3872)$, one would expect its observables to be utterly dominated by long-distance interactions.  This contradiction, in part, has led to the long-standing view that $\chi_{c1}(3872)$ contains at least some component of the fundamental charmonium state $\chi_{c1}(2P)$~\cite{Suzuki:2005ha}.  But an alternate short-range, color-attractive configuration is available to the $\chi_{c1}(3872)$, in the form of a diquark-antidiquark pair: $(cu)_{\mathbf{\bar 3}} (\bar c \bar u)_{\mathbf{3}}$.

In fact, one approach using this paradigm, the \textit{dynamical diquark model}~\cite{Brodsky:2014xia,Lebed:2017min}, has made strides in successfully representing the $\chi_{c1}(3872)$ as an exotic diquark-antidiquark state, as well generating the full accompanying spectra of both tetraquark and pentaquark exotic multiplets in multiple flavor sectors~\cite{Lebed:2017min,Giron:2019bcs,Giron:2019cfc,Giron:2020fvd,Giron:2020qpb,Giron:2020wpx,Giron:2021fnl,Giron:2021sla}.  These advances include the incorporation of effects such as spin- and isospin-dependent interactions, SU(3$)_{\rm{flavor}}$ mixing, and most recently, mixing between diquark-antidiquark states and nearby di-hadron  thresholds~\cite{Lebed:2022vks}.

While the original dynamical diquark model calculations were performed assuming that di-hadron thresholds close in mass to those of the diquark-antidiquark states can be neglected---which imposes the framework of the Born-Oppenheimer (BO) approximation---the incorporation of di-hadron threshold mixing can be accomplished through its rigorous generalization;  this so-called \textit{diabatic formalism} was originally developed for, and has long been used in, molecular physics~\cite{Baer:2006}.   First introduced into hadronic physics by Ref.~\cite{Bruschini:2020voj} to analyze exotic states produced by the mixing of heavy quarkonium $Q \bar Q$ with di-hadron thresholds, the diabatic framework also provides a method through which diquark-antidiquark states mixing with di-hadron thresholds can be analyzed~\cite{Lebed:2022vks}.

Almost all exotic states lie above the energy threshold of the lowest possible open-heavy-flavor hadron pair with the same $J^{PC}$ and flavor quantum numbers.  While these states may, in some cases, be approximated as bound states (which is the assumption of Refs.~\cite{Lebed:2022vks,Bruschini:2020voj,Bruschini:2021cty}), the more accurate treatment is to view these states as resonant poles within scattering processes.  The unification of the diabatic formalism with scattering theory, again using $Q \bar Q$/di-hadron mixing, was pioneered in Ref.~\cite{Bruschini:2021ckr}.  Here, we expand upon the work of Ref.~\cite{Lebed:2022vks} by developing the same techniques for $\de$-$\bde$/di-hadron mixing.

\begin{table} \label{tab:exotics}
\caption{Examples of heavy-quark exotic candidates lying particularly close in mass ($< 15$~MeV) to a di-hadron threshold.}
\begin{tabular}{cc}
\hline \hline
    Exotic candidate & Di-hadron threshold  \rule{0pt}{2.6ex} \\
    \hline \rule{0pt}{2.6ex}
    $\chi_{c1}(3872)$ & $D^0 \bar D^{*0}$ \\
    $Z_c(3900)$ & $D \bar D^*$ \\
    $Z_c(4020)$ & $D^* \! \bar D^*$ \\
    $P_c(4312)$ & $\Sigma_c \bar D$ \\
    $P_c(4450)/P_c(4457)$ & $\Sigma_c \bar D^*$ \\
    $Z_b(10610)$ & $B \bar B^*$ \\
    $Z_b(10650)$ & $B^* \! \bar B^*$ \\
    \hline \hline 
\end{tabular}
\end{table}

This paper is organized as follows.  In Sec.~\ref{sec:DynDiqModel} we define the features of the dynamical diquark model, which generates the spectrum of heavy-quark exotic hadrons studied here.  Section~\ref{sec:diabatic} describes the \textit{diabatic} formalism that generalizes the adiabatic formalism inherent in the BO approximation used by the original dynamical diquark model.  The diabatic formalism is incorporated in Sec.~\ref{sec:Scatt} into scattering theory, particularly in order to study open-flavor heavy-meson elastic scattering processes, in which exotic resonances (ultimately originating as dynamical-diquark states) may occur.  In Sec.~\ref{sec:results}, we first reprise our previous bound-state calculations, and then present numerical results for hidden-charm scattering cross sections and discuss the diverse interesting features that arise.  Section~\ref{sec:Concl} summarizes our conclusions and indicates the next directions for research.

\section{The Dynamical Diquark Model} \label{sec:DynDiqModel}

The dynamical diquark \textit{picture}~\cite{Brodsky:2014xia} provides key context for the construction of the full scattering model developed in this paper.  In the original picture, quark pairs ($qQ$) and ($\bar q \bar Q$) in (attractive) color-triplet configurations ($Q$ being heavy) are produced within relative proximity of each other, and with a high relative momentum with respect to the opposite pair; such a scenario occurs in an appreciable fraction of $Q\bar Q$ production processes.  Thus, the diquarks $\de \equiv (qQ)_{\bar{\bf 3}}$ and $\bde \equiv (\bar q \bar Q)_{\bf 3}$ can naturally form as compact objects, especially since heavy $Q$ have less Fermi motion.  Due to confinement, $\de$ and $\bde$ remain bound to each other via a color flux tube.  The kinetic energy associated with the high relative momentum is then converted into the potential energy of the flux tube as the distance between the diquarks increases, the $\de$-$\bde$ separation eventually reaching a maximum as the relative momentum between the compact diquarks drops toward zero.  With an appreciable distance now separating the quark-antiquark pairs that can form color singlets, this configuration has difficulty hadronizing, allowing it to persist long enough to be observed as an exotic tetraquark resonance.  The analogous process for the pentaquark case~\cite{Lebed:2015tna} can also be described using this mechanism by substituting $\bde \rightarrow \bar \theta $, where the color-triplet {\it triquark\/} is defined by $\bar \theta \equiv [ \bar Q (q_1 q_2)_{\bar {\bf 3}} ]^{\vphantom\dagger}_{\bf 3}$.

The dynamical diquark \textit{model} is then constructed from this picture by implementing the BO approximation for QCD, as described in detail in Sec.~\ref{sec:diabatic}\@.  This approximation, which has been extensively used to study heavy hybrid mesons, provides the most natural formalism for describing such a quasi-static system.  The end result of applying the BO approximation is the generation of a set of effective static potentials, which in turn are used to produce a full spectrum of state multiplets.  These \textit{BO potentials} may be explicitly calculated on the lattice (see, {\it e.g.},  Refs.~\cite{Juge:1999ie,Juge:2002br,Capitani:2018rox}).  The multiplets of states within these potentials are denoted by a set of five quantum numbers: $\Lambda^\epsilon_\eta (nL)$, where $\Lambda^\epsilon_\eta$ define the BO potentials through the symmetries of the light degrees of freedom (d.o.f.), and $n,L$ indicate the familiar radial and angular momentum quantum numbers defining the orbitals of each BO potential.  Explicitly, the labels $\Lambda^\epsilon_\eta$ designate irreducible representations of the group $D_{\infty h}$, which describes the symmetries inherent to a cylinder whose axis coincides with the characteristic radial separation vector $\mathbf{\hat r}$ of the heavy quasiparticle pair.  

A more detailed discussion of these potentials, as well as their application to $\de \bde$ and $\de \bar \theta$ systems, may be found in Refs.~\cite{Lebed:2017min,Giron:2019bcs}.  For the purpose of this analysis, Refs.~\cite{Giron:2019bcs,Giron:2020fvd,Giron:2020qpb} are especially important by  providing clear numerical indications that these potentials correctly describe multiplet mass averages for heavy-quark exotic states in each light-flavor sector ({\it e.g.}, $c\bar c q\bar q^\prime$ in $1S$ and $1P$ states, $c\bar c qqq$, $c\bar c s\bar s$, $b\bar b q\bar q^\prime$), and these multiplets are shown to accommodate the $J^{PC}$ quantum numbers of all known exotics.  The multiplet mass averages may then be resolved into a fine-structure spectrum by introducing Hamiltonian spin- and isospin-dependent operators that are expected to be the ones most relevant for describing the fine-structure effects.  In general, the number of free parameters in the model is then $(n+1)$, the coefficients of the $n$ fine-structure operators included in the analysis, plus the diquark (triquark) mass $m_{\de}~(m_{\theta})$.  A phenomenological fixing of these parameters, where one fits to the numerical value of each so that the best-understood exotic states emerge naturally, is the approach of Refs.~\cite{Giron:2019bcs,Giron:2019cfc,Giron:2020fvd,Giron:2020qpb,Giron:2020wpx,Giron:2021sla,Giron:2021fnl}; a mass prediction for every member of the complete spectrum of states then immediately follows.

\section{The Diabatic Approach}\label{sec:diabatic}

The incorporation of the diabatic approach into the dynamical diquark model~\cite{Lebed:2022vks} signifies a departure from the strict framework of the BO approximation to its rigorous generalization~\cite{Baer:2006}, and we reprise its development for hadronic systems here. To describe a (nonrelativistic) system consisting of two heavy color sources interacting through light (quark and gluon) fields, one begins with the Hamiltonian
\begin{equation} 
\label{eq:SepHam}
H=K_{\rm heavy} + H_{\rm light} =
\frac{\mathbf{p}^2}{2 \mu_{\rm heavy}} + H_{\rm light},
\end{equation}
where $H_{\rm light}$ contains the light-field static energy, as well as the heavy-light interaction.  Under the BO framework, one writes the solutions to the corresponding \schro equation as 
\begin{equation} 
\label{eq:AdExp}
|\psi \rangle = \sum_{i} \int d\mathbf{r} \, \tilde \psi_i
(\mathbf{r}) \, |\mathbf{r} \rangle \:
|\xi_i(\mathbf{r}) \rangle,
\end{equation}
where $|\mathbf{r} \rangle$ are defined as states of heavy source pairs with separation vector $\mathbf{r}$, and $|\xi_i(\mathbf{r}) \rangle$ is the $i^{\rm th}$ eigenstate of $H_{ \rm light}$.  Note that the heavy and light states here reference the same value of $\mathbf{r}$; Eq.~(\ref{eq:AdExp}) is called the \textit{adiabatic expansion}, although the expression at this point remains general.  The set $\{ |\xi_i(\mathbf{r}) \rangle \}$ forms a complete, orthonormal basis for the light d.o.f.\ at any given $\mathbf{r}$, but in general, configuration mixing occurs at different values of $\mathbf{r}$:  $\langle \xi_j (\mathbf{r}') | \xi_i (\mathbf{r^{\vphantom\prime}}) \rangle \neq 0$ even for $j \neq i$.  Inserting Eq.~(\ref{eq:AdExp}) into the \schro equation and taking inner products with $\langle \xi_j(\mathbf{r})|$, after some manipulations one arrives at 
\begin{equation}
\sum_i \left( - \frac{\hbar^2}{2 \mu_{Q \bar Q}} [\mathbf{\nabla} + \tau (\mathbf{r})]^2_{ji} + [V_j (\mathbf{r}) - E] \, \delta_{ji}  \right) \! \tilde \psi_i (\mathbf{r}) = 0,
\end{equation}
where the functions $\tau (\mathbf{r})_{ji}$, known as \textit{Non-Adiabatic Coupling Terms} (NACTs), are defined as 
\begin{equation}
\mathbf{\tau}_{ji}(\mathbf{r}) \equiv \langle \xi_j (\mathbf{r}) | \nabla \xi_i (\mathbf{r}) \rangle.
\end{equation}
If, in addition, the heavy d.o.f.'s are sufficiently heavy compared to the light d.o.f.'s, then one may approximate the light d.o.f.'s as instantaneously (\textit{adiabatically}) adapting to changes in the heavy-source separation, which in this notation reads $\langle \xi_i (\mathbf{r}') | \xi_i (\mathbf{r^{\vphantom\prime}} ) \rangle \approx 1$ for small changes $\mathbf{r}' \neq \mathbf{r}$, the \textit{adiabatic approximation}.  Additionally, at values of $\mathbf{r}',\mathbf{r}$ where the light-field eigenstates do not appreciably mix, one has $\langle \xi_j (\mathbf{r}') | \xi_i (\mathbf{r}) \rangle \approx 0$ for $j \neq i$, which is called the \textit{single-channel approximation}.  These two approximations define the full \textit{BO approximation}, and are conveniently summarized by the single condition on the NACTs:
\begin{equation} \label{eq:NACT}
\tau_{ji} (\mathbf{r}) = \langle \xi_j (\mathbf{r}) | \nabla \xi_i (\mathbf{r}) \rangle \approx 0.
\end{equation}
For systems containing a heavy (hence static) $Q\bar Q$ pair, unquenched lattice-QCD calculations have long found that this approximation works well in regions far from energy thresholds for on-shell di-meson production.  Close to these thresholds, the static light-field energies experience an avoided level-crossing, thus demonstrating the explicit breaking of the single-channel approximation~\cite{Bali:2005fu,Bulava:2019iut}.   In order to discuss more general mixed states that may have such energies, one may adopt the rigorous generalization of the BO approximation known as the \textit{diabatic formalism}~\cite{Baer:2006}.  This method rewrites the expansion of the solution Eq.~(\ref{eq:AdExp}) as 
\begin{equation} 
\label{eq:DiaExp}
|\psi \rangle = \sum_{i} \int d\mathbf{r}' \tilde \psi_i
(\mathbf{r}' \! , \mathbf{r}_0) \: |\mathbf{r}' \rangle \:
|\xi_i(\mathbf{r}_0) \rangle,
\end{equation}
where $\mathbf{r}_0$ is a free parameter.  Here again, the completeness of the basis $\{ |\xi_i(\mathbf{r}) \rangle \}$, regardless of the choice of $\mathbf{r}$, is crucial. In analogy to the previous procedure, one inserts the expansion Eq.~(\ref{eq:DiaExp}) into the \schro equation and takes inner products with $\langle \xi_j(\mathbf{r}_0) |$, thus producing
\begin{equation}\label{eq:DiaSchro}
\sum_{i} \left[ - \frac{\hbar^2}{2 \mu_{i}} \de_{ij}  \nabla ^2 +
V_{ji}(\mathbf{r,r_0})-E \de_{ji} \right] \! \tilde \psi_i (\mathbf{r,r_0}) = 0.
\end{equation}

Now the object of interest is $V_{ji}$, which is known as the \textit{diabatic potential matrix}; it is defined as 
\begin{equation}
V_{ji}(\mathbf{r,r_0}) \equiv \langle \xi_j (\mathbf{r}_0)|
H_{\rm light} |\xi_i(\mathbf{r}_0) \rangle.
\end{equation}
The NACT method and the diabatic-potential method are rigorously equivalent, as shown in Refs.~\cite{Bruschini:2020voj,Baer:2006}, but the latter is more convenient for our numerical simulations.  As discussed in Ref.~\cite{Bruschini:2020voj}, one may choose ${\bf r}_0$ far from potential-energy level crossings, such that the  states $|\xi_i(\mathbf{r}_0) \rangle$ may be unambiguously identified with pure, unmixed configurations.  For the specific application to dynamical-diquark states with a fixed value of ${\bf r}_0$, we identify the diagonal elements of this matrix as the static light-field energies $V_{\de \bde}$ associated with a pure $\de \bde$ state and its corresponding di-meson thresholds $V_{M_1 \bbar M_2}^{(i)}$, $i = 1, 2, \ldots , N$.  Explicitly, $V_{ji}$ may then be written as 
\begin{equation} \label{eq:FullV}
\text V=
\begin{pmatrix}
V_{\de \bde}(\mathbf{r}) & V_{\rm mix}^{(1)}(\mathbf{r})  & \cdots &
V_{\rm mix \vphantom{\bbar M_2}}^{(N)}(\mathbf{r}) \\
V_{\rm mix}^{(1)}(\mathbf{r}) & 
V_{M_1 \bbar M_2}^{(1)}(\mathbf r) &
&
\\
\vdots
& & \ddots \\
V_{\rm mix \vphantom{\bbar M_2}}^{(N)}(\mathbf{r}) & & &
V_{M_1 \bbar M_2}^{(N)}(\mathbf r) \\
\end{pmatrix},
\end{equation}
where we ignore direct mixing terms between any two di-meson configurations ({\it i.e.}, the suppressed elements are zero).  For the purposes of this work, we set each pure di-meson energy to be the free energy of the state, {\it i.e.},
\begin{equation}
V_{M_1 \bbar M_2}^{(i)}(\mathbf r) \to T_{M_1 \bbar M_2} = M_1 + M_2 \, .
\end{equation}

One could of course instead replace $V_{M_1 \bbar M_2}^{(i)}(\mathbf r)$ with a mildly attractive potential ({\it e.g.}, pion-exchange interactions or the effects of triangle singularities), as suggested in Ref.~\cite{Lebed:2022vks}.

\section{Scattering Theory} \label{sec:Scatt}

As noted in the Introduction, the diabatic formalism provides a method to study mixed but still formally bound states.  In contrast, nearly all of the exotic candidates have been observed solely through their strong-interaction decays, and therefore should properly be treated as resonances in scattering theory, {\it i.e.}, as poles in a scattering $S$-matrix.

Here we review the construction of the $K$-matrix formalism as a method of retrieving the $S$-matrix for coupled-channel eigenstates of the \schro equation, specifically using the method of Ref.~\cite{johnson1973generalized}.  The $K$-matrix has several advantages over the $S$-matrix, in particular that it can be chosen to be real and symmetric (assuming time-reversal symmetry), and that pole terms induced by distinct resonances, even heavily overlapping ones, may be simply added together in the $K$-matrix (unlike for the $S$-matrix).  In this work, we consider only elastic scattering of asymptotically pure di-meson configurations.  As discussed in Ref.~\cite{Bruschini:2021cty}, this type of scattering, mediated by the short-range mixing of di-meson and $\de \bde$ states, is the natural physical process in which to study the asymptotic behavior of solutions to Eq.~(\ref{eq:DiaSchro}).  Collecting the set of linearly independent solutions to the \schro equation into a matrix $\Psi$, one may write the asymptotic behavior as 
\begin{equation}
\Psi(\mathbf{r}) = \mathbf{J}(\mathbf{r}) - \mathbf{N}(\mathbf{r}) \mathbf{K},
\label{eq:KmatrixScatter}
\end{equation}
where $\mathbf{K}$ denotes the $K$- (or \textit{reaction}) matrix, and $\mathbf{J}$ and $\mathbf{N}$ are the (diagonal) solutions to the \schro equation in the $r \to \infty$ limit, at which only the centrifugal part of the potential remains significant.  Following Ref.~\cite{johnson1973generalized}, we choose the closed-channel elements (channels with thresholds above the total energy $E$) of both matrices to be proportional to their corresponding modified spherical Bessel functions $i_{\ell_j}, k_{\ell_j}$ ($x_i \equiv r k_i$, where $k_i$ is the wave number for the $i^{\rm th}$ channel):
\begin{eqnarray}
\rm{J}_{ij} & = & x_i \! \cdot \! i_{\ell_j}(x_i) \, \delta_{ij}, \nonumber \\
\rm{N}_{ij} & = & x_i \! \cdot \! k_{\ell_j}(x_i) \, \delta_{ij},
\label{eq:JNclosed}
\end{eqnarray}
while the open-channel elements (channels with thresholds below the total energy $E$) are set to be the Riccati-Bessel functions,
\begin{eqnarray}
\rm{J}_{ij} & = & x_i \! \cdot \! j_{\ell_j}(x_i) \, \delta_{ij}, \nonumber \\
\rm{N}_{ij} & = & x_i \! \cdot \! n_{\ell_j}(x_i) \, \delta_{ij} .
\label{eq:JNopen}
\end{eqnarray}
Formally, one may then write $\mathbf{K}$ as a function of $\mathbf{J}$, $\mathbf{N}$, and the log-derivative $\mathbf y$ of the matrix solution $\Psi$, $\mathbf y \equiv \Psi ' \Psi ^{-1}$:
\begin{equation}\label{eq:Kmatrix}
\mathbf{K} = \left( \mathbf{y} \mathbf{N} - \mathbf{N}'\right)^{-1} \left( \mathbf{y} \mathbf{J} - \mathbf{J}' \right).
\end{equation}
In the sign convention for $\textbf{K}$ imposed by Eq.~(\ref{eq:KmatrixScatter}) (see Ref.~\cite{Morrison:2007} for alternate sign conventions for all of these quantities), the $\textbf{S}$-matrix is obtained as: 
\begin{equation}\label{eq:Smatrix}
\mathbf{S} = (\mathbf{ I} - i \mathbf{K}_{\rm oo})^{-1}(\mathbf{I} + i \mathbf{K}_{\rm oo}),
\end{equation}
where $\mathbf{K}_{\rm oo}$ denotes the sub-matrix of $\mathbf{K}$ containing \textit{only} elements that connect open channels to other open channels.  That Eq.~(\ref{eq:Smatrix}) can be expressed solely in terms of $\mathbf K_{\rm oo}$ relies directly upon the specific forms of Eqs.~(\ref{eq:JNclosed})--(\ref{eq:JNopen}), as is thoroughly explained in Ref.~\cite{PhysRevA.32.1241}.  Reference~\cite{johnson1973generalized} also provides a method for numerically calculating Eq.~(\ref{eq:Kmatrix}) using the reduced Numerov method, which has already been employed extensively for  solving dynamical-diquark \schro equations (starting with Ref.~\cite{Giron:2019bcs}).

We now briefly comment on the form of the solutions contained in $\Psi$.  Since this analysis is concerned only with the elastic scattering of asymptotically pure di-meson states, we restrict this discussion to the elements of $\Psi$ associated with those states.  With $V_{M_1 \bbar M_2}(\mathbf{r})=T_{M_1 \bbar M_2}$, the well-known \textit{unmixed} solutions are:
\begin{equation}
\psi^{(i)}_{J^{PC} \! , m_J}(\mathbf{r}) = \sqrt{\frac{2}{\pi} \mu^{(i)} p^{(i)}} \, i^{\ell^{(i)}_k} j_{\ell^{(i)}_k} (p^{(i)} r ) {\rm Y}^{J,m_J}_{\ell^{(i)}_k s^{(i)}_k}(\mathbf{\hat r}),
\end{equation}
where ${\rm Y}^{J,m_J}_{\ell^{(i)}_k s^{(i)}_k}$ are irreducible tensors of rank $J$,
\begin{equation}
\text{Y}^{J,m_J}_{\ell^{(i)}_k s^{(i)}_k} \equiv \langle \mathbf{\hat r} | \ell,s,J,m_J \rangle = \sum_{m_\ell,m_s} C^{m_\ell,m_s,m_J}_{\ell,s,J} Y^{m_\ell}_\ell (\mathbf {\hat r} ) \, \xi_s^{m_s},
\end{equation}
built with the conventional Clebsch-Gordan coefficients $C^{m_\ell,m_s,m_J}_{\ell,s,J}$, spherical harmonics $Y^{m_\ell}_\ell (\mathbf {\hat r} )$, and spinors $\xi_s^{m_s}$. In addition,
\textit{k} and ($i$) denote the $k^{\rm th}$ partial wave of the $i^{\rm th}$ di-meson threshold with quantum numbers $J^{PC}$, while $j_\ell$ is the $\ell^{\rm th}$ spherical Bessel function of the first kind, $p^{(i)} = \sqrt{2\mu^{(i)} ( E-T^{(i)})}$ is the relative momentum (or wave number) of the di-meson pair, and $\sqrt{\frac{2}{\pi} \mu^{(i)} p^{(i)}}$ is a factor introduced in Ref.~\cite{Bruschini:2021ckr} to normalize the full solution in terms of energy $E$:
\begin{equation}
    \langle \Psi_{E'} | \Psi_{E} \rangle = \delta ( E' - E). 
\end{equation}
One may go further by using the large-argument asymptotic expression for spherical Bessel functions,
\begin{equation}
j_{\ell}(pr) \to \frac{1}{pr} \text{sin} \left( pr - \ell \frac{\pi}{2} \right).
\end{equation}
This form allows for \textit{mixed} solutions to be clearly expressed using well-known elastic scattering theory ({\it e.g.}, Eq.~(11.17) in Ref.~\cite{taylor2012scattering}): The effect of mixing with a short-range attractive state, in this case $\de \bde$, enters as a channel- and momentum-dependent phase shift $\delta$ in the unmixed asymptotic wavefunctions of the di-meson configurations.  Explicitly,
\begin{multline}
\frac{1}{p^{(i)}r} \text{sin} \left( p^{(i)}r - \ell \frac{\pi}{2} \right) \longrightarrow \\ e^{i \delta^{(i)}_{\ell} } \! \frac{1}{p^{(i)}r} \text{sin} \left( p^{(i)}r - \ell \frac{\pi}{2} + \delta^{(i)}_{\ell} \right).
\end{multline}
Summing over all partial waves $k$ (and adopting the notation of Ref.~\cite{Bruschini:2021ckr} as closely as possible), we have
\begin{multline}
\psi^{(i)}_{J^{PC} \! , m_J}(\mathbf{r}) = \frac{1}{r} \sqrt{\frac{2 \mu^{(i)}}{ \pi p^{(i)} }  }  \sum_{k} i^{\ell^{(i)}_k} \! a^{(i)}_{J^{PC} \! ;k} \\ \times \frac{1}{p^{(i)}r} \text{sin} \left( p^{(i)}r - \ell_k \frac{\pi}{2} + \delta^{(i)}_{J^{PC} \! ; k} \right) {\rm Y}^{J,m_J}_{\ell^{(i)}_k s^{(i)}_k}(\mathbf{\hat r}) ,
\end{multline}
where the usual scattering coefficients $a^{(i)}_{J^{PC} \! ;k}$ keep track of the weighted amplitude that each partial wave contributes to the overall $J^{PC}$ state.  Finally, one may now write the asymptotic wavefunction of a di-meson to di-meson scattering state ($ i \leftarrow i' $), in specific partial waves $k'\equiv (\ell^{(i')} , s^{(i')}) \to k\equiv (\ell^{(i)} , s^{(i)})$, as 
\begin{multline}
\psi^{i \leftarrow i'}_{J^{PC} \! ,m_J;k'} (\mathbf{r}) \mathbf{=} \frac{1}{r} \sqrt{\frac{2 \mu^{(i)}}{ \pi p^{(i)} }  }  \sum_{k} i^{\ell^{(i)}_k} \times \\  \left[  \delta_{i i'} \delta_{k k'} \text{sin} \left( p^{(i)}r - \ell_k^{(i)}\frac{\pi}{2} \right) + p^{(i)} f^{i \leftarrow i'}_{J^{PC} \! ; k,k'} e^{i ( p^{(i)} r - \ell^{(i)}_{k} \frac{\pi}{2} )}  \right] \\ \times \text{Y}^{J,m_J}_{k} (\mathbf{\hat r}),
\end{multline}
with $f^{i \leftarrow i'}_{J^{PC} ; k,k'}$ being the partial-wave scattering amplitude.  In the present analysis, $f^{i \leftarrow i'}_{J^{PC} ; k,k'}$ are the objects of interest, since one may extract the elastic-scattering cross sections directly from these scattering amplitudes.

We do so, again following the work of Ref.~\cite{Bruschini:2021ckr}, and thus provide a proof-of-concept calculation of elastic-scattering cross sections for the di-meson configurations (mediated by coupling to $\de \bde$ states) as discussed in Sec.~\ref{sec:diabatic}\@.  This may be done using the $S$-matrix by calculating the scattering amplitude 
\begin{equation} \label{eq:ScatterAmp}
f^{i \leftarrow i'}_{J^{PC}} = \frac{ ( S - \mathbb{I} )_{i i'} }{2ip^{(i)}},
\end{equation}
with which one may calculate the $J^{PC}$-specific partial cross section 
\begin{equation}\label{eq:sigma}
\sigma^{i \leftarrow i'}_{J^{PC}} = \frac{4\pi(2J+1)}{(2s_{M^{(i')}_1} +1)(2s_{\bbar M^{(i')}_2} + 1)}   \sum_{k,k'} |  f^{i \leftarrow i'}_{J^{PC};k,k'} |^2 . 
\end{equation}
For the purposes of this calculation, we instead calculate a normalized cross section $\bar \sigma$~\cite{Bruschini:2021ckr},
\begin{equation}\label{eq:sigmabar}
\bar \sigma_{J^{PC}}^{i \leftarrow i'} = \sum_{k,k'} | p^{(i)}  f^{i \leftarrow i'}_{J^{PC};k,k'} |^2 ,
\end{equation}
which allows for a clearer investigation of the behavior near threshold (where phase space, and hence $\sigma$, vanishes), as well as providing more unequivocal indications of resonant behavior, in which fully saturated resonances are expected to reach the maximum allowed value of unity for $\bar \sigma$. 

\section{Results} \label{sec:results}

In this analysis, we assume the mixing elements of the diabatic-potential matrix in Eq.~(\ref{eq:FullV}) to have the simple Gaussian form~\cite{Bruschini:2020voj}:
\begin{equation} \label{eq:Mixpot}
|V_{\rm mix}^{(i)} (r)| = \frac{\Delta}{2}
\exp \! \left\{ -\frac 1 2 \frac{\left[
V^{\vphantom\dagger}_{\de \bde}(r) -
T_{M_1 \bbar M_2 }^{(i)} \right]^2}{\Lambda^2} \right\} ,
\end{equation}
where $\Delta$ is the strength of the mixing and $\Lambda$ is a width parameter, both with units of energy.  To produce meaningful results, $\Delta$ must be large enough to induce sufficient mixing with $\de \bde$ states that clearly indicates the importance of nearby di-meson thresholds, while $\Lambda$ must be small enough not to induce excess mixing with thresholds far from the original $\de \bde$ state; until lattice-QCD simulations are able to provide specific values for these parameters, their magnitudes remain constrained only by these qualitative constraints.  One may rewrite $\Lambda$ as 
\begin{equation}
\Lambda = \rho \sigma,
\end{equation}
where $\rho$ may be identified as the radial scale of the mixing, while $\sigma$ is the \textit{string tension} of the $\de \bde$ configuration.  As discussed in Ref.~\cite{Bruschini:2020voj}, this particular form of the mixing potential, which is motivated by results of lattice QCD~\cite{Bulava:2019iut}, acts as a phenomenological placeholder, in anticipation of future precision lattice simulations.  This mixing potential is also a different creature than the one used in the original diabatic works such as Ref.~\cite{Bruschini:2020voj}; in the original calculations it refers to $Q\bar Q \to (Q \bar q)(\bar Q q)$ string breaking, while in the present calculations it refers to the rearrangement interaction $(Qq)(\bar Q \bar q) \to (Q \bar q)(\bar Q q)$.

In fact, if one truly wishes to rigorously perform the calculations of this work with the intention of accurate comparison to experiment, then the form of the mixing potential will likely be more complicated.  A complete treatment should include every fundamental channel (both open \textit{and} closed flavor) in the diabatic potential matrix of Eq.~(\ref{eq:FullV}), transitions between these configurations must be considered, and in order to couple to these channels to allow realistic decays, the mixing potential must must allow for more complicated forms than a simple universal Gaussian function.  Additionally, in contrast to the work of Ref.~\cite{Bruschini:2020voj}, the mixing potential connecting $\de \bde$ states to meson-meson thresholds may have strong correlations with the particular spin state of the diquarks, and such a dependence should be included in some form as well.

With these caveats in mind, we start by reproducing the results of the bound-state formalism in Ref.~\cite{Lebed:2022vks}, with a slight variation of model parameters:
\begin{eqnarray}
\rho & = & 0.165 ~\rm{fm}, \\
\Delta & = & 0.295 ~\rm{GeV}, \\
m_{c \bar q}  =  m_{q \bar c} & = & 1927.1~\rm{MeV}, \label{eq:cqmass} \\ 
m_{c \bar s} =  m_{s \bar c} & = & 1944.6 ~\rm{MeV}, \label{eq:csmass}
\end{eqnarray}
and, for the ground-state BO potential $\Sigma^+_g$,
\begin{equation}
\label{eq:sgmapot}
V_{\de \bde}(r)=- \frac{\alpha}{r} + \sigma r + V_0 + m_{\de} +
m_{\bde} \, ,
\end{equation}
where $\alpha,\sigma,$ and $V_0$ are $0.053 ~\text{GeV$\cdot$fm},
1.097 ~\text{GeV/fm},$ and $-0.380 ~\text{GeV}$, respectively~\cite{Berwein:2015vca}.  Hence, $\Lambda$ in Eq.~(\ref{eq:Mixpot}) is 0.181~GeV\@.  We note that applying the hybrid $Q\bar Q$ potential inputs obtained from lattice simulations for the $\de \bde$ case is reasonable, since both are color ${\bf 3}$-$\bar{\bf 3}$ potentials between two heavy sources.  For BO potentials above $\Sigma_g^+$, which generally tend to mix with each other, the extension of this formalism is straightforward.  However, in this work we focus solely on the $\Sigma_g^+$ potential, since all exotics found to date appear to be accommodated within its orbitals~\cite{Giron:2019bcs}.  These results are presented in Tables~\ref{tab:ccqq}, \ref{tab:ccss}, and \ref{tab:ccqs}.  As in Ref.~\cite{Lebed:2022vks}, the mixing parameters are retrieved by fitting to the $\chi_{c1}(3872)$ mass central value $3871.65$~MeV reported by the PDG~\cite{ParticleDataGroup:2022pth}, while keeping the same diquark mass $m_{c \bar q}$ [Eq.~(\ref{eq:cqmass})] found in Ref.~\cite{Giron:2021sla}.  Additionally, the mixing parameters are moderately constrained to reproduce certain behaviors of the mixing angle between the $\de \bde$ and $D \bar D^*$ components: {\it i.e.}, the mixing angle $\theta (r)$ must smoothly and quickly vary between 0 and $\pi / 2$ as $r$ decreases/increases away from the \textit{critical radius} $r_c$, which is defined as the separation for which $V_{\de \bde}(r_c)$ equals the $D \bar D^*$ threshold mass (see Refs.~\cite{Bruschini:2020voj,Lebed:2022vks}).  Again, we note that these mixing parameters $\rho, \Delta$ are not uniquely defined by this fit, and thus only serve as working values for the present analysis.  With these inputs, the diquark mass $m_{c \bar s}$ is then fixed [Eq.~(\ref{eq:csmass})] by requiring the $0^{++}~\ccss$ state to have mass equal to the central value 3921.7~MeV for $X(3915)$ given by the PDG~\cite{ParticleDataGroup:2022pth}.

Once these parameters are fixed, the diabatic dynam\-ical-diquark model Hamiltonian (not yet including fine structure) for each tetraquark flavor sector, $\ccqq,~c \bar c q \bar s,$ and $c \bar c s \bar s$, is completely specified.  This assertion, of course, assumes that the mixing parameters are universal, and not unique to each threshold or flavor sector.  Some work towards this end, specifically to include heavy-quark spin-symmetry breaking effects, has been carried out in Ref.~\cite{bruschini2023heavyquark}, where the author calculates transition rates between the elementary state (in that case, $Q \bar Q$) and its corresponding thresholds. The primary result of this work is a demonstration of how to handle the threshold nonuniversality that occurs between different di-meson thresholds ({\it e.g.}, $D \bar D^*$ {\it vs.} $D^* \bar D^*$), which constitutes one direction in which one may move past the universality assumption of our mixing potential.  Using the formalism described in Sec.~\ref{sec:Scatt}, we may then directly produce flavor- and $J^{PC}$-specific cross sections as functions of center-of-mass energy.  In aggregate, these results are presented in Figs.~\ref{fig:ccqq1++cc}--\ref{fig:ccqs2+}.  Some universal characteristics include the stability of all major functional features in $\bar \sigma$ [Eq.~(\ref{eq:sigmabar})] upon minor variations of the phenomenologically determined parameters $\rho$, $\Delta$, and $m_{\delta}$.  Additionally, we find resonant behavior to occur in all but one of the cross sections, which is consistent with the calculations performed under the bound-state framework of Refs.~\cite{Bruschini:2020voj,Lebed:2022vks}.  That is, we find resonances in the near proximity of all predicted bound states.

\subsection{$\ccqq$}

\renewcommand{\arraystretch}{1.2}
\begin{table*}[ht]
\caption{Calculated eigenvalues and component-state admixtures for
the $\ccqq$ sector obtained from solving Eq.~(\ref{eq:DiaSchro}) for
specific $J^{PC}$ numbers.  Suppressed entries indicate contributions that are
individually finite but $< \! 1\%$, or that give no contribution.}
\setlength{\tabcolsep}{9pt}
\renewcommand{\arraystretch}{1.2}
\begin{tabular}{c  c  c c c c c  } 
 \hline\hline
 $J^{PC}$ & $E \ ({\rm MeV})$ & $\de \bde$ & $D \bar D^*$ &
 $D_s \bar D_s$ & $D^* \bar D^*$  & $D_s^* \bar D_s^*$  \\
 \hline
 $0^{++}$ & 3903.83  &  69.8\% & &  22.7\%  &
  6.9\% &  \\ 
 $1^{++}$ & 3871.65  &  9.1\% &  90.9\% &  & &   \\
 $2^{++}$ & 3917.44  &  86.0\% & &  1.5\% &
  10.4\% & 1.5\%  \\
  \hline 
 & & &  $D \bar D_1$ & $D \bar D_2^*$ & $D^* \bar D_1$
  \\
  \hline 
$1^{--}$ & 4269.58 & 44.0\% & 51.2\% & 2.4\% & 1.5\%
  \\
 \hline\hline
\end{tabular}
\label{tab:ccqq}
\end{table*}

Unique amongst the $\ccqq$ sector are the $1^{++}$ results presented in Fig.~\ref{fig:ccqq1++cc}. Here, in addition to $\de \bde$, we incorporate a $c \bar c$ channel [representing the fundamental $\chi_{c1}(2P)$ state] into the diabatic-potential matrix, with a mixing potential connecting this channel to the (same) corresponding di-meson thresholds.  This particular simulation, unlike others in the $\ccqq$ category, necessarily produces only isosinglet amplitudes.  The mixing, for which we adopt the same form as that for $\de \bde$-$M_1 \bbar M_2$, is parameterized using the results of Ref.~\cite{Bruschini:2020voj}: specifically, $\rho_{c \bar c} = 0.3 ~\rm{fm}$ and $\Delta_{c \bar c} = 0.130 ~\rm{GeV}$.  A direct comparison with Fig.~\ref{fig:ccqq1++}, in which the $c \bar c$ channel is removed, reveals that its inclusion in Fig.~\ref{fig:ccqq1++cc} can result in the formation of a secondary peak containing significant overlap with the peak appearing at the mass of $\chi_{c1}(3872)$.  Additional contributions from the processes $D^* \bar D^*$ and $D_s \bar D^*_s$ are also induced by the inclusion of the $\chi_{c1}(2P)$ component in Fig.~\ref{fig:ccqq1++cc}.  We also note the appearance of threshold effects at the $D_s \bar D^*_s$ mass ($\sim$4081~MeV) in Figs.~\labelcref{fig:ccqq1++,fig:ccqq1++cc}, since this model makes no attempt to address the Okubo-Zweig-Iizuka (OZI) suppression required for transitions between states containing $q\bar q$ and $s\bar s$.  Lastly, we find  (but do not exhibit here) that the usual elastic scattering phase $\delta_\ell$, defined by
\begin{equation}
S^{ii'}_\ell \equiv e^{2i\delta^{i \leftarrow i'}_\ell},
\end{equation}
exhibits resonant behavior (sharp transitions in $\delta$ reaching above $\pi / 2$) at both major peaks in Fig.~\ref{fig:ccqq1++cc}.  However, this conclusion is only true for the mixed S-D partial wave, {\it i.e.}, $D \bar D^* (\ell=0) \leftrightarrow D \bar D^* (\ell=2)$.  The pure S-wave process, $D \bar D^* (\ell=0) \leftrightarrow D \bar D^* (\ell=0)$, \textit{nearly} reaches $\pi / 2$ at the $\chi_{c1}(3872)$ mass, and then smoothly trails off at higher energies; but if one varies the parameters $\rho$ or $\Delta$, then it is possible to induce a value of $\delta$ that rises above $\pi / 2$ in the pure S-wave process at the same mass.  Under this variation, the same resonant behavior as in the other partial-wave channels is still observed.  This result implies that the current framework can easily accommodate a pair of resonant states $\de \bde$ and $\chi_{c1}(2P)$, either fully overlapping or clearly discernable, each mixing with nearby di-hadron thresholds.  The results of Fig.~\ref{fig:ccqq1++} may also be used to extract the corresponding decay width for the $D \bar D^*$ channel if one converts the data back to the physical cross section of Eq.~(\ref{eq:sigma}).  We find 0.4~MeV for the width of the peak, which may be compared to the PDG value $0.44\pm0.13$~MeV for $\chi_{c1}(3872)$ decaying to $D^0 \bar D^{*0}$~\cite{ParticleDataGroup:2022pth}.  We note that our result is found through extrapolation, since the full peak structure is cut off by threshold itself.  Although this assumption must be taken with some caution, the closeness of these two values implies that it is straightforward to find values of $\rho$ and $\Delta$ that exactly accommodate exactly both the correct mass and width of $\chi_{c1}(3872)$.

In fact, the case of $\chi_{c1}(3872)$ is particularly interesting, because it indicates limitations on the freedom to choose diabatic couplings.  While we have noted the stability of basic morphological features of $\bar \sigma$ under variations of the parameters $\rho$, $\Delta$, and $m_\delta$, the fact that the precise mass and width of $\chi_{c1}(3872)$ are now highly constrained means that the values of the diabatic parameters, given a particular functional form such as in Eq.~(\ref{eq:Mixpot}), must be carefully chosen in order to maintain agreement with experiment.  These adjustments must be revisited as additional channels [{\it e.g.}, $J/\psi \, \pi^+ \pi^-$ for $\chi_{c1}(3872)$] and spin- and isospin-dependent couplings in the Hamiltonian are incorporated into future iterations of these calculations.  Even so, certain features such as the large $D \bar D^*$ content and small but significant $\de \bde$ content of $\chi_{c1}(3872)$ should remain robust.

Our $0^{++}$ results in Fig.~\ref{fig:ccqq0++} show a wide, fully saturated peak at 3900~MeV in the process $D \bar D \rightarrow D \bar D$, with nontrivial modifications from both $D_s \bar D_s$ and $D^* \bar D^*$ thresholds as well.  This result is consistent with expectations inferred from Table~\ref{tab:ccqq}, where the bound-state approximation produces significant contributions from the corresponding thresholds to a state with matching energy, 3903.83~MeV\@.  In Fig.~\ref{fig:ccqq0++}, the impacts of threshold effects in the lineshapes are clearly visible.

Conversely, for $2^{++}$ scattering (Fig.~\ref{fig:ccqq2++}), we observe a sharp peak in $D \bar D$ near 3910~MeV, which can be unambiguously assigned to the corresponding state (3917.44~MeV) of Table~\ref{tab:ccqq}\@.  Outside of this peak in the $2^{++}$ cross section, there are relatively small contributions in all but the  $D^* \bar D^*$ channel.  We also observe the same preferential $2^{++}$ coupling to $D^* \bar D^*$ in Table~\ref{tab:ccqq}, despite its threshold ($\sim$4014~MeV) being significantly higher in mass than the $D_s \bar D_s$ threshold ($\sim$3937~MeV).  In Ref.~\cite{Lebed:2022vks}, this enhancement is attributed to the fact that the $D^* \bar D^*$ threshold coupling to $2^{++}$ allows an S-wave coupling, which is naturally expected to dominate over $\ell>0$ configurations (D-wave for $D_{(s)} \bar D_{(s)}$ in $2^{++}$) in scattering processes.

While the sharpness of the peak in Fig.~\ref{fig:ccqq2++} suggests the existence of a clear $\de\bde$ resonance with $J^{PC} = 2^{++}$ that should be immediately detectable by experiment, it is important to remind the reader that these widths arise from calculations using incomplete physical information.  A more detailed treatment of the threshold couplings and mixing potential, as discussed at the beginning of Sec.~\ref{sec:results}, is essential before the widths may be compared with experiment.

For example, in the present case, the isoscalar $2^{++}$ channel is already known to feature the $c\bar c$ candidate $\chi_{c2}(2P)$ at $3922.5 \pm 1.0$~MeV~\cite{ParticleDataGroup:2022pth} (which could certainly have been included in this analysis, in the same manner as done in Fig.~\ref{fig:ccqq1++cc}), and this state has a substantial width of about 35~MeV, likely largely due to its observed (D-wave) $D\bar D$ decay mode.  A comparison between the calculation of widths through conventional methods ({\it i.e.}, as performed in Ref.~\cite{Bruschini:2021cty} for $c\bar c$ states in the diabatic formalism) with those obtained from the scattering formalism will appear in future work.

In this work we now include bound-state results (Table~\ref{tab:ccqq}) for the $1^{--}$ $c\bar c q\bar q^\prime$ channel (which did not appear in the results of Ref.~\cite{Lebed:2022vks}), and also present the corresponding cross section (Fig.~\ref{fig:ccqq1--}).  The energy interval (4.15--4.50~GeV) exhibited for the analysis in this channel is restricted to impose stringent requirements upon which thresholds to include, in order to admit only those expected to generate the most physically significant effects.  Thus, we include only thresholds for meson pairs with relatively small individual widths ($< 50$~MeV), and (with the exception of $D_s^* \bar D_s^*$) that couple to $1^{--}$ in an S-wave.  This calculation produces a resonant peak with an extraordinarily small width (only 4.2~MeV), but again we caution the reader that the widths of states appearing in these plots are based upon incomplete physical input.  At a mass of about 4240~MeV, this peak is clearly sensitive to the $D_s^* \bar D_s^*$ threshold, which again requires an OZI-suppressed amplitude to couple to $c\bar c q\bar q$.  It is natural to identify this peak with $\psi(4230)$, even though this state's open-charm decay modes are poorly known (only $\pi^+ D^0 D^{*-}$ has thus far been seen~\cite{ParticleDataGroup:2022pth}).  We also note a nearly 30-MeV shift of the resonant peak from the bound-state energy predicted by the corresponding state in Table~\ref{tab:ccqq}.  While we have yet to explicitly calculate the expected bound-state mass shifts that arise from the perturbative introduction of couplings to open thresholds, Ref.~\cite{Bruschini:2021cty} provides a rough estimate of what might be expected through their analogous calculation in $c\bar c$-$D^{(*)}_{(s)} \bar D^{(*)}_{(s)}$ mixing.  A comparison to the largest shift noted in that work, roughly $28$~MeV, allows for the reasonable identification of the peak in Fig.~\ref{fig:ccqq1--} with the $1^{--}$ bound state of Table~\ref{tab:ccqq}.  Beyond this peak, the $1^{--}$ channel as displayed in Fig.~\ref{fig:ccqq1--} exhibits an abundance of threshold behaviors in all presented cross sections.

\subsection{$c \bar c s \bar s$ and $c \bar c q \bar s$}

The full suite of $c \bar c s \bar s$ results is presented in Figs.~\ref{fig:ccss0++}--\ref{fig:ccss2++}, while the $c \bar c q \bar s$ (or $c \bar c s \bar q$) results appear in Figs.~\ref{fig:ccqs1+}--\ref{fig:ccqs2+}.  Beginning with our $0^{++}$ findings for the $c \bar c s \bar s$ sector (Fig.~\ref{fig:ccss0++}), we observe further agreement with our bound-state predictions (Table~\ref{tab:ccss}) in the appearance of a fully saturated peak at 3920~MeV in $D \bar D \rightarrow D \bar D$.  One may note the similarity of this lineshape with the analogous one in the $\ccqq$ sector (Fig.~\ref{fig:ccqq0++}).  Such results are a direct result of the fact that this formalism is currently ``blind'' to any effects due to strangeness, other than through explicit differences in diquark and meson masses.  We expect this effect to diminish as additional SU$(3)_{\text{flavor}}$ symmetry breaking is incorporated.

In the $1^{++}$ results for this sector (Fig.~\ref{fig:ccss1++}), we find a relatively wide peak centered at 3925~MeV in the $D \bar D^* \rightarrow D \bar D^*$ cross section.  While this result may appear to discourage assignment to the $1^{++}$ state in Table~\ref{tab:ccss} (3968.47~MeV), we note the relatively long tail present in this peak structure, and also recall the up-to-30~MeV downwards shift that may be caused by the introduction of open thresholds.  These two facts argue that an assignment of the peak in Fig.~\ref{fig:ccss1++} to the $1^{++}$ bound state in Table~\ref{tab:ccss} is not unreasonable, and indeed, show how strong threshold effects can be in certain channels.  As threshold structures are abundant throughout the full results of this analysis, we draw attention to their absence in both hidden-flavor $1^{++}$ resonances [Figs.~\labelcref{fig:ccqq1++,fig:ccss1++}] at the $D^* \bar D^*$ threshold.  As symmetry forbids an S-wave $1^- 1^- \to 1^{++}$ coupling, this threshold has only a D-wave coupling to $1^{++}$. Thus, these results provide further evidence for the dominance of S-wave couplings in scattering processes.

Lastly, we find the  $2^{++}$-channel scattering (Fig.~\ref{fig:ccss2++}) to yield a sharp (but not fully saturated) peak around 3925~MeV, which falls within the aforementioned 30-MeV interval for reasonable identification with the corresponding bound state of Table~\ref{tab:ccss}\@.  In Fig.~\ref{fig:ccss2++}, we observe a uniquely interesting case, in which the state appears to be dragged below the previously open threshold of $D_s \bar D_s$ [although Table~\ref{tab:ccss} disallows admixture to this state because the bound state (3949.33~MeV) was found to lie above the $D_s \bar D_s$ threshold ($\sim$3937~MeV)].  In the scattering context, $D_s \bar D_s$ only couples to $2^{++}$ through a D-wave, and therefore is still expected to be suppressed compared to S-waves.

\renewcommand{\arraystretch}{1.2}
\begin{table*}[ht!]
\caption{The same as in Table~\ref{tab:ccqq}, for the $\ccss$ sector.}
\begin{tabular}{c  c  c c c c c  }
\hline\hline
\rule{0pt}{2.6ex}
$J^{PC}$ & $E \ ({\rm MeV})$ &  $\de \bde$ & $D_s \bar D_s$ &
$D^* \! \bar D^*$ & $D_s \bar D_s^*$ & $D_s^* \bar D_s^*$  \\
\hline
$0^{++}$ & 3921.69 &  55.7\% &  35.4\% &
 7.1\% & & 1.2\%  \\
\hline
$1^{++}$ & 3968.47 &  90.4\% & &  1.2\% &
 7.8\% &   \\
\hline
$2^{++}$ & 3949.33 &  82.1\% & & 15.5\% &
 & 2.1\%  \\
\hline\hline
\end{tabular}
\label{tab:ccss}
\end{table*}

The $c \bar c q \bar s$ sector provides another opportunity to examine the nearly unbroken SU(3)$_{\rm flavor}$ symmetry present in this calculation.  A near-perfect overlap is observed for $1^{+}$ elastic $D^* \bar{D}_s$ and $D \bar{D}^*_s$ scattering processes (Fig.~\ref{fig:ccqs1+}).  We find no resonant behavior in these results, consistent with the $1^+$ prediction of Table~\ref{tab:ccqs}, which indicates an eigenstate (3912.73~MeV) below the lowest available di-meson threshold ($\sim$3975~MeV).  Additionally, we find fully saturated peaks in both the $0^{+}$ (Fig.~\ref{fig:ccqs0+}) and $2^{+}$ (Fig.~\ref{fig:ccqs2+}) results, centered just above and just below 3950~MeV, respectively.  Of the two, the peak found in $2^{+}$ $D \bar D_s \rightarrow D \bar D_s$ notably has the smallest apparent width of any appearing in this analysis (but with the same caveats discussed above).  In both cases, the location of the peak differs only slightly from the predictions of Table~\ref{tab:ccqs}, which, interestingly, our calculations show can be attributed to the introduction of the $D \bar D_s$ threshold ($\sim$3833~MeV), which lies well below the predicted eigenvalues. One may also contrast the contributions of the $D^* \bar D_s$ and $D \bar D_s^*$ processes in Figs.~\labelcref{fig:ccqs1+,fig:ccqs2+}.  We see that a bound-state calculation in which over $90\%$ of the content is $\de \bde$ (\textit{i.e.}, Fig.~\ref{fig:ccqs2+} but not Fig.~\ref{fig:ccqs1+}) produces no obvious structure in $\bar \sigma$ for scattering processes with thresholds far above the resonance.  This conclusion is corroborated by the results of Figs.~\labelcref{fig:ccss1++,fig:ccqs0+}\@.

In addition, the inputs in this sector are completely fixed by the phenomenological fits to the other flavor sectors, and thus provide useful benchmarks for comparison against experiment.  The $1^+$ state of Table~\ref{tab:ccqs} (3912.73 MeV) in particular, which is generally unaffected by the changes introduced in the present calculation, may ultimately be associated with the observed $Z_{cs}(3985)$~\cite{ParticleDataGroup:2022pth}, once multiplet fine-structure effects are included, especially the mixing of strange states in distinct $1^+$ SU(3)$_{\rm flavor}$ multiplets~\cite{Giron:2021sla}.  This assignment works especially well when one compares the admixtures of the Table~\ref{tab:ccqs} state with the fact that $Z_{cs}(3985)$ has been observed as a $D_s \bar D^* + D_s^* \bar D$ resonance~\cite{Ablikim:2020hsk}.  The difference between these two masses ($\sim$70~MeV) is well within the largest fine-structure mass-splitting effect predicted for diquark-antidiquark states in this sector~\cite{Giron:2021sla}.

An additional comparison is available from the $1^{++}$ state of Table~\ref{tab:ccss}: Although the mass difference is much larger [$\sim$170~MeV, corresponding to the $\ccss$ candidate $\chi_{c1}(4140)$], it is not yet known how the fine structure of \textit{diabatic} dynamical diquark states differs from that of states that are blind to threshold effects, particularly once effects sensitive to the larger strange-quark mass are properly included. 

\renewcommand{\arraystretch}{1.2}
\begin{table*}[ht!]
\caption{The same as in Tables~\ref{tab:ccqq} \& \ref{tab:ccss}, for
the $c\bar c q\bar s$ sector.}
\begin{tabular}{ c  c  c c c c   }
\hline\hline
\rule{0pt}{2.6ex} 
$J^P$ & $E \ ({\rm MeV})$ & 
$\de \bde$ & $D^* \! \bar D_s$ & $D \bar D_s^*$ & $D^* \! \bar D_s^*$
 \\
\hline
$0^+$ & 3969.04 &  95.2\% & & &  4.5\%  \\
\hline
$1^+$ & 3912.73 &  71.9\% &  13.7\%  &
 13.4\% & \\
\hline
$2^+$ & 3951.42 &  92.5\% &  1.5\% &
 1.5\% &  4.5\%  \\
\hline\hline
\end{tabular}
\label{tab:ccqs}
\end{table*}

\section{Conclusions} \label{sec:Concl}

We have reviewed the incorporation of the diabatic formalism, a rigorous extension of the well-known Born-Oppenheimer approximation that is designed to include effects due to the presence of two-particle thresholds, into the dynamical diquark model.  While our previous work addresses states formed in the immediate vicinity of these thresholds (the bound-state approximation), this paper develops a scattering framework capable of describing not only exotic states lying close to such thresholds, but also those that lie quite far from them (and thus have no obvious interpretation as a di-hadron molecular state).

Using the bound-state approximation, we first reproduce our previous flavor- and $J^{PC}$-specific calculations of energy eigenvalues and fractions of both diquark-antidiquark and di-meson components within the corresponding eigenstates.  We then summarize the construction of the K-matrix formalism as a method to retrieve the S-matrix, in order to calculate asymptotic scattering amplitudes of coupled-channel, elastic meson-meson collision processes (the most natural ones in which to study resonance and threshold behaviors).  We validate the physical expectation that asymptotically free meson-meson pairs develop resonance structures through their short-range interaction with diquark-antidiquark channels.  These scattering amplitudes are calculated numerically for the hidden-charm system (with zero, hidden, and open strangeness), and then are directly used to produce all corresponding CM energy-dependent cross sections, which comprise the main results of this work.

We confirm the expected resonant behavior in all flavor- and $J^{PC}$-specific cross sections, and also observe several instances of threshold-induced structures such as cusp effects.  In addition, the peak of every resonance is calculated to occur not far from the energy of its corresponding bound-state eigenvalue.  We observe shifts of these resonances down from the bound-state energies once the couplings to open thresholds are included, in agreement with expectations that thresholds are generally ``attractive.''  While nearly all of these resonant behaviors reach the maximum value allowed by unitarity, some prominent examples reach as low as $\sim 75\%$ of this value.

Although this analysis is mostly limited to meson-meson scattering coupled to diquark-antidiquark channels described by the dynamical diquark model, we find evidence that the conventional $c \bar c$ state $\chi_{c1}(2P)$ may be incorporated separately into the $c\bar c q\bar q$ $1^{++}$ channel, producing two resonant components that may overlap to form $\chi_{c1} (3872)$.    In general, a complete calculation would include all diquark-antidiquark and $c\bar c$ states in every allowed $J^{PC}$ channel.

While these results are quite promising, they do not yet distinguish explicit spin- and isospin-multiplet members.  The incorporation of such fine-structure analysis has been accomplished for multiple flavor sectors in the original (adiabatic) dynamical diquark model, and thus will be straightforward to include in its diabatic form; this extension will be one major thrust of future work.  In addition, this analysis does not incorporate SU(3$)_{\rm flavor}$ symmetry-breaking effects beyond explicit differences in the diquark masses $m_{cq}$ and $m_{cs}$, and in meson masses $m_{D_{\vphantom{s}}^{(*)}}, m_{D_s^{(*)}}$, \textit{etc}.  Such additional effects, not to mention OZI suppression, are expected to have substantial impact on the scattering processes discussed here.  Lastly, the widths of the resonances implied by these cross-section plots are not always suitable for direct comparison with experiment, as they are calculated using a universal, and hence, incomplete set of couplings to meson-meson thresholds, as well as (aside from the one example in Fig.~\ref{fig:ccqq1++cc}) lacking couplings to closed-flavor channels.  Thus, future work will also use well-known techniques to calculate physical strong-decay widths and shifts of energy eigenvalues due to open-threshold di-meson pairs that lie well below the diabatically mixed eigenstates studied here---{\it i.e.}, the pairs that represent their physical decay channels.

\vspace{1em}

\begin{acknowledgments}
This work was supported by the National Science Foundation (NSF) under 
Grants No.\ PHY-1803912 and PHY-2110278.
\end{acknowledgments}

\begin{widetext}

\begin{figure}
    \centering
    \includegraphics[scale=0.65]{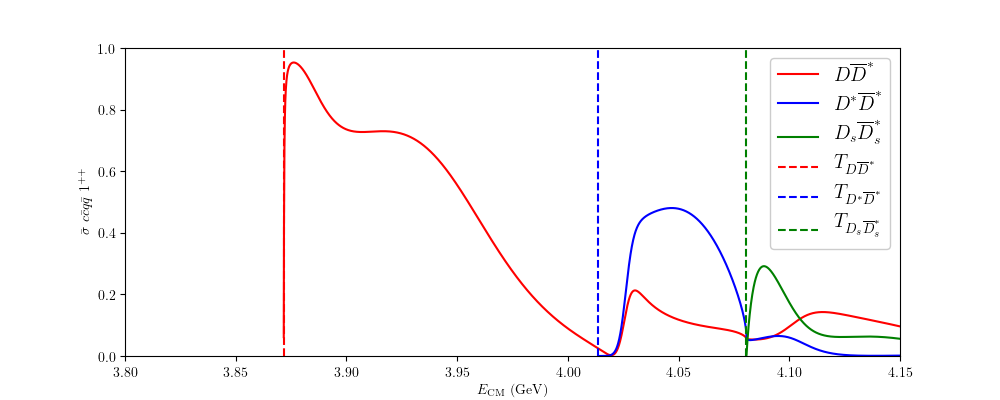}
    \caption{The dimensionless cross sections $\bar \sigma$ of Eq.~(\ref{eq:sigmabar}) (solid lines) for elastic open-charm $D^{(*)}_{(s)} \bar D^{(*)}_{(s)}$ scattering processes as functions of center-of-momentum frame energy $E_{\rm CM}$.  The $\bar \sigma$ curves are presented in the same order as that of increasing mass for their corresponding thresholds $T$ (dashed lines).  This figure presents results for the flavor content $c\bar c q\bar q$ in the isosinglet channel with $J^{PC} = 1^{++}$, and includes a contribution from the conventional $c\bar c$ state $\chi_{c1}(2P)$.}
    \label{fig:ccqq1++cc}
\end{figure}

\begin{figure}
    \centering
    \includegraphics[scale=0.65]{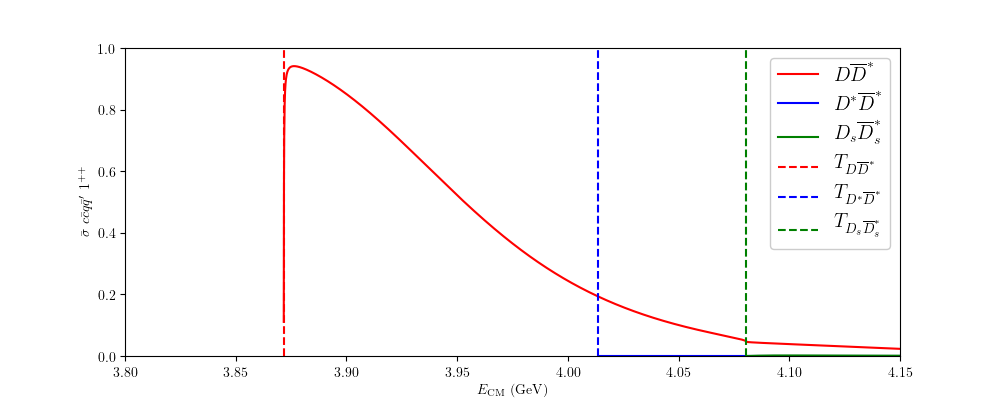}
    \caption{The same as in Fig.~\ref{fig:ccqq1++cc}, except suppressing the $\chi_{c1}(2P)$ contribution, and (if the small $D_s \bar D_s^*$ contributions are also suppressed) not necessarily limited to the isoscalar combination of $\ccqq$.}
    \label{fig:ccqq1++}
\end{figure}

\begin{figure}
    \centering
    \includegraphics[scale=0.65]{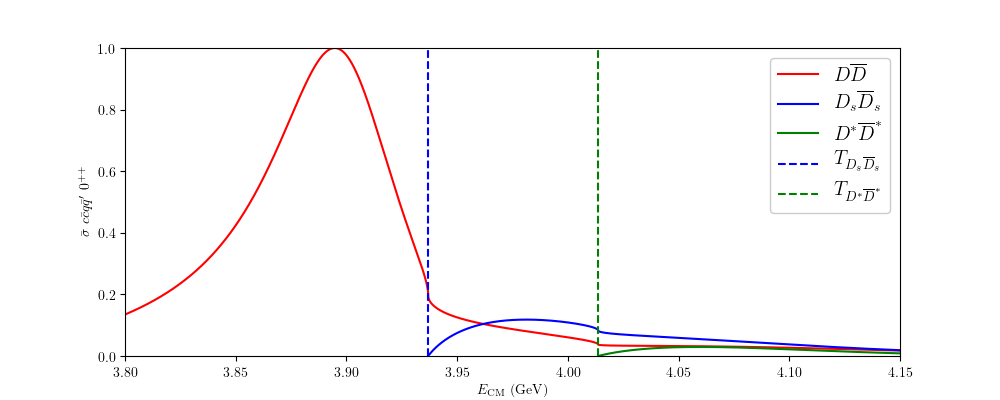}
    \caption{The same as in Fig.~\ref{fig:ccqq1++}, for elastic open-charm $\ccqq$ $D^{(*)}_{(s)} \bar D^{(*)}_{(s)}$ scattering processes with $J^{PC}=0^{++}$.}
    \label{fig:ccqq0++}
\end{figure}

\begin{figure}
    \centering
    \includegraphics[scale=0.65]{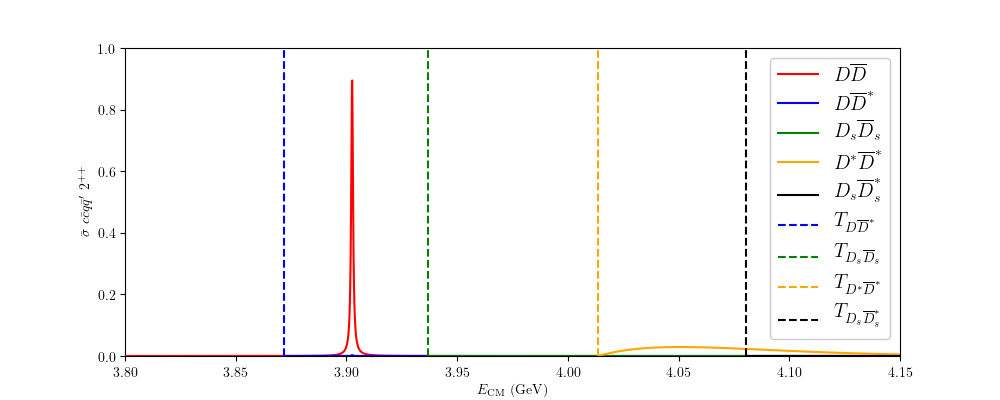}
    \caption{The same as in Fig.~\ref{fig:ccqq1++}, for elastic open-charm $\ccqq$ $D^{(*)}_{(s)} \bar D^{(*)}_{(s)}$ scattering processes with $J^{PC}=2^{++}$.}
    \label{fig:ccqq2++}
\end{figure}

\begin{figure}
    \centering
    \includegraphics[scale=0.65]{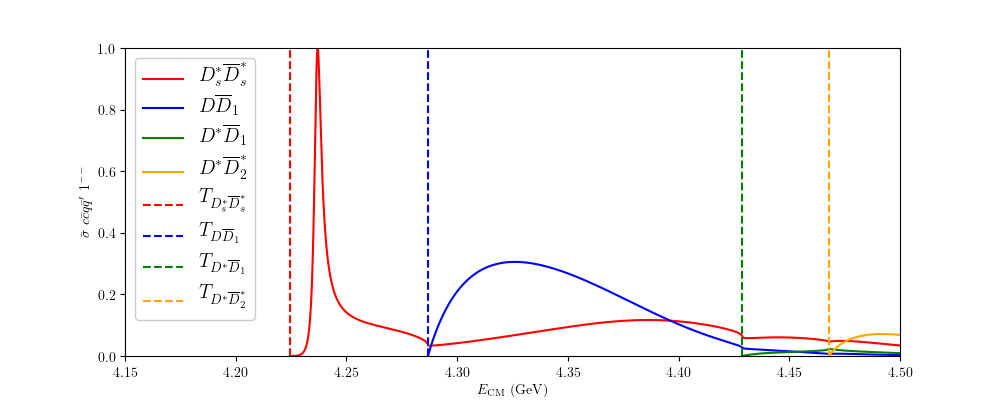}
    \caption{The same as in Fig.~\ref{fig:ccqq1++}, for elastic open-charm $\ccqq$ scattering processes with $J^{PC}=1^{--}$.  The energy range has been adjusted to 4.15--4.50~GeV in order to capture the relevant behavior of the cross section.  The thresholds are all S-wave, except for $D_s^* \bar D_s^*$, which is $P$-wave.}
    \label{fig:ccqq1--}
\end{figure}

\begin{figure}
    \centering
    \includegraphics[scale=0.65]{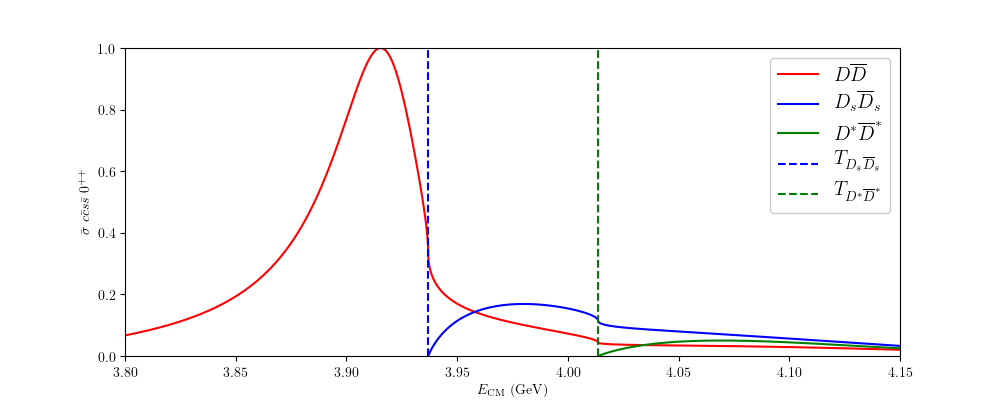}
    \caption{The same as in Fig.~\ref{fig:ccqq0++}, but now for the $c\bar c s\bar s$ channel $J^{PC} = 0^{++}$.}
    \label{fig:ccss0++}
\end{figure}

\begin{figure}
    \centering
    \includegraphics[scale=0.65]{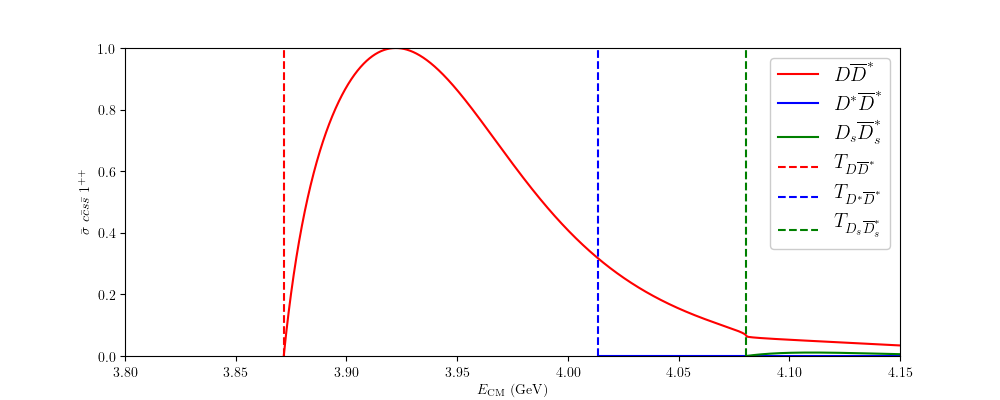}
    \caption{The same as in Fig.~\ref{fig:ccss0++}, for the $c\bar c s\bar s$ channel $J^{PC} = 1^{++}$.}
    \label{fig:ccss1++}
\end{figure}

\begin{figure}
    \centering
    \includegraphics[scale=0.65]{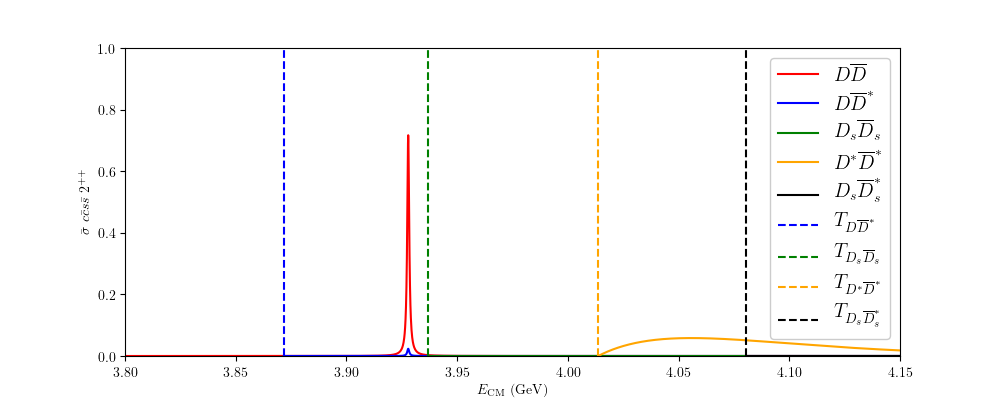}
    \caption{The same as in Fig.~\ref{fig:ccss0++}, for the $c\bar c s\bar s$ channel $J^{PC} = 2^{++}$.}
    \label{fig:ccss2++}
\end{figure}

\begin{figure}
    \centering
    \includegraphics[scale=0.65]{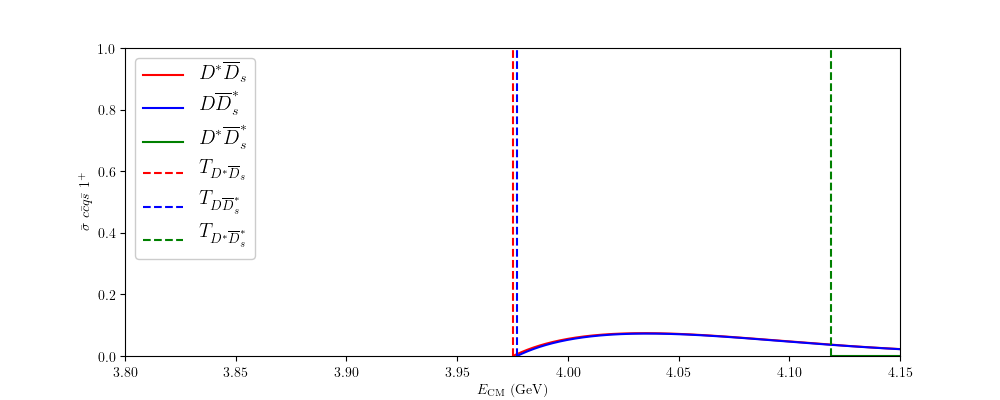}
    \caption{The same as in Fig.~\ref{fig:ccqq1++}, but now for the $c\bar c q\bar s$ channel $J^P = 1^+$.}
    \label{fig:ccqs1+}
\end{figure}

\begin{figure}
    \centering
    \includegraphics[scale=0.65]{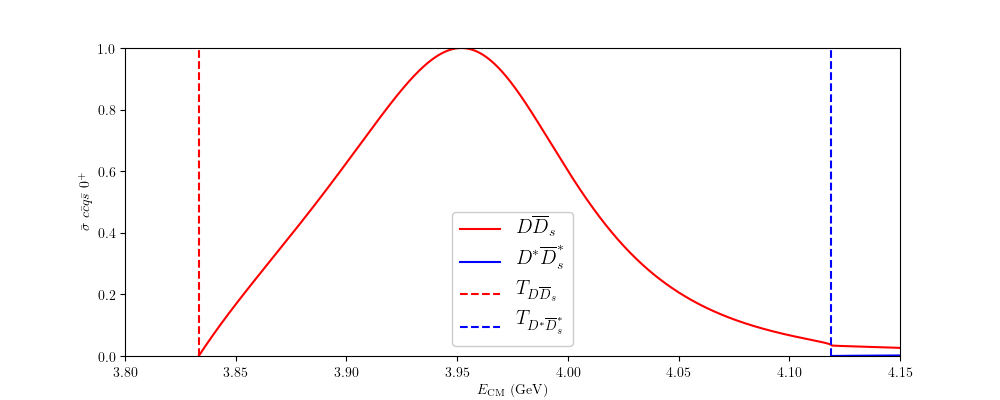}
    \caption{The same as in Fig.~\ref{fig:ccqs1+}, for the $c\bar c q\bar s$ channel $J^P = 0^+$.}
    \label{fig:ccqs0+}
\end{figure}

\begin{figure}
    \centering
    \includegraphics[scale=0.65]{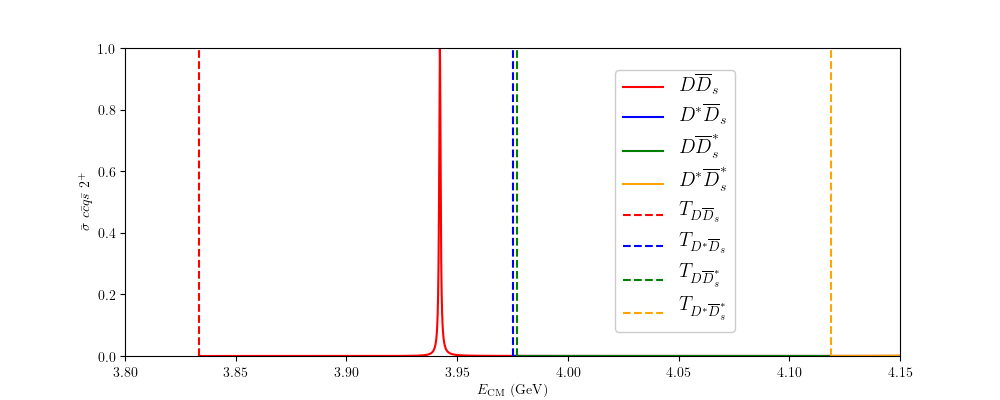}
    \caption{The same as in Fig.~\ref{fig:ccqs1+}, for the $c\bar c q\bar s$ channel $J^P = 2^+$.}
    \label{fig:ccqs2+}
\end{figure}

\end{widetext}
\bibliographystyle{apsrev4-2}
\bibliography{diquark}
\end{document}